\documentclass[aps,prl,superscriptaddress,floatfix,showpacs,twocolumn]{revtex4}

\usepackage{verbatim}

\usepackage{amsmath}
\usepackage{amsfonts}
\usepackage{amssymb}
\usepackage{graphicx}
\usepackage{bm}
\usepackage{color}
\usepackage{epsf}
\usepackage[hypertex]{hyperref}

\usepackage{psfrag}

\def\bea{\begin{eqnarray}}
\def\eea{\end{eqnarray}}
\def\beas{\begin{eqnarray*}}
\def\eeas{\end{eqnarray*}}
\def\beqas{\begin{eqnarray*}}
\def\eqas{\end{eqnarray*}}
\def\beq{\begin{equation}}
\def\eeq{\end{equation}}
\def\beqd{\begin{displaymath}}
\def\eeqd{\end{displaymath}}
\def\eqd{\end{displaymath}}

\def\slashchar#1{\setbox0=\hbox{$#1$}
   \dimen0=\wd0
   \setbox1=\hbox{/} \dimen1=\wd1
   \ifdim\dimen0>\dimen1
      \rlap{\hbox to \dimen0{\hfil/\hfil}}
      #1
   \else\begin{eqnarray}
      \rlap{\hbox to \dimen1{\hfil$#1$\hfil}}
      /
   \fi}

\begin{document}
\title
{Vector meson electroproduction at next-to-leading order }
\author{D.Yu~Ivanov}
\affiliation{Sobolev Institute of Mathematics, 630090 Novosibirsk, Russia}
\author{ L.~Szymanowski}
\affiliation{Soltan Institute for Nuclear Studies, Hoza 69, 00-681
Warsaw, Poland}
\author{ G. Krasnokov }
\affiliation{Department of Theoretical Physics, St.Petersburg State University,
198904,
St. Petersburg, Russia}


\begin{abstract}

\noindent
The process of a light neutral vector meson electroproduction is studied in
the framework of QCD factorization
in which the amplitude factorizes in a convolution of the nonperturbative
meson distribution amplitude  and the generalized parton densities
with the perturbatively calculable hard-scattering amplitudes.
We derive a complete set of hard-scattering amplitudes at next-to-leading
order (NLO) for the production
of vector mesons, $V=\rho^0, \omega, \phi$.
\end{abstract}

\maketitle

\noindent
{\bf 1.}
The process of
elastic
neutral vector meson electroproduction on a nucleon,
\begin{equation}
\gamma^*(q)\, N(p) \to V(q^\prime)\, N(p^\prime)
\, ,
\label{process}
\end{equation}
where $V=\rho^0,\,  \omega, \, \phi $, was studied in fix target
and in HERA collider experiments.
The primary motivation for the strong interest in this process
(and in the similar process of heavy quarkonium production) is that
it can potentially serve to constrain the gluon density in a nucleon
\cite{Ryskin:1992ui,BFGMS94}.
On the theoretical side, the large negative virtuality of the photon,
$q^2=-Q^2$, provides a hard
scale for the process which justifies the application of QCD factorization
methods that allow to separate the contributions to the amplitude
coming from different scales.
The factorization theorem \cite{CFS96} states that in a scaling limit,
$Q^2\to \infty$ and $x_{Bj}=Q^2/2(p\cdot q)$ fixed, a vector meson is
produced
in the longitudinally polarized state by the longitudinally polarized
photon
and that the amplitude of the
process (\ref{process}) is given by a convolution of the
nonperturbative
meson distribution amplitude (DA) and the generalized parton densities
(GPDs)
with the perturbatively calculable hard-scattering amplitudes.
In this contribution we present the results of our calculation of the
hard-scattering amplitudes at NLO.

\noindent{\bf 2.}
$p^2=p^{\prime 2}=m_N^2$ and $q^{\prime \, 2}=m^2_M$,
where $m_N$ and $m_M$ are a proton mass and a meson mass
respectively.
The invariant c.m. energy $s_{\gamma^*p}=(q+p)^2=W^2$. We define
\begin{eqnarray}
&&
\Delta=p^\prime -p \, , \ \ P=\frac{p+p^\prime}{2} \, , \ \ t=\Delta^2 \, ,
\nonumber \\
&&
(q-\Delta )^2=m^2_M \, , \ \ x_{Bj} =\frac{Q^2}{W^2+Q^2} \, .
\label{not1}
\end{eqnarray}

We introduce two light-cone vectors
\begin{equation}
n_{+}^2=n_{-}^2=0 \, , \ \ n_+ n_- = 1 \, .
\label{not2}
\end{equation}
Any vector $a$ is decomposed as
\begin{equation}
a^\mu=a^+n_+^\mu+a^-n_-^\mu+a_\perp \, , \  a^2=2\, a^+a^- - \vec a^2 \, .
\label{not3}
\end{equation}
We choose the light-cone
vectors in such a way that
\begin{eqnarray}
&&
p=(1+\xi)W n_+ + \frac{m_N^2}{2(1+\xi)W}\, n_- \, ,
\label{not33} \\
&&
p^\prime=(1-\xi)W n_+
+\frac{(m_N^2+\vec \Delta^2)}{2(1-\xi)W}\, n_- +\Delta_\perp \, .
\nonumber
\end{eqnarray}
We are interested in the kinematic region where
the invariant transferred momentum, $t$,
is small, much smaller than $Q^2$.
In the scaling limit variable $\xi$ which parametrizes the
plus component of the momentum transfer equals $\xi=x_{Bj}/(2-x_{Bj} )$.

GPDs are defined as the matrix element of the light-cone quark
and gluon operators \cite{Diehl:2003ny}:
\begin{eqnarray}
&&
F^q (x,\xi,t)=
\nonumber \\
&&
=\frac{1}{2}\! \int\! \frac{d\lambda}{2\pi}
e^{i x (P z)}\langle
p^\prime |\bar q \! \left(-\frac{z}{2}\right)\!\! \not \! n_-
q\! \left(\frac{z}{2}\right)\!
|p\rangle|_{z=\lambda n_-}  
\nonumber \\
&&
=\frac{1}{2(Pn_- )}\biggl[
{\cal H}^q (x,\xi,t)\, \bar u(p^\prime)\not \! n_- u(p)
\nonumber \\
&&
\left.
+
{\cal E}^q (x,\xi,t)\, \bar
u(p^\prime)\frac{i\sigma^{\alpha\beta}n_{-\alpha}\Delta_\beta}{2\,m_N}
u(p)
\right] \, , 
\label{qGPD}
\end{eqnarray}
\begin{eqnarray}
&&
F^g (x,\xi,t)=\frac{1}{(Pn_-)}\int\frac{d\lambda}{2\pi}
e^{i x (P z)}\,
n_{-\alpha}n_{-\beta}
\nonumber \\
&&
\langle
p^\prime |G^{\alpha\mu} \left(-\frac{z}{2}\right)
G^{\beta}_\mu \left(\frac{z}{2}\right)
|p\rangle|_{z=\lambda n_-}
\nonumber \\
&&
\nonumber
=\frac{1}{2(Pn_- )}\Bigl[
{\cal H}^g (x,\xi,t)\, \bar u(p^\prime)\not \! n_- u(p)
\nonumber \\
&&
+
{\cal E}^g (x,\xi,t)\, \bar
u(p^\prime)\frac{i\sigma^{\alpha\beta}n_{-\alpha}\Delta_\beta}{2\,m_N}
u(p)
\Bigr] \, .
\label{gGPD}
\end{eqnarray}
In both cases the insertion of the path-ordered gauge factor between the
field operators is implied.
Momentum fraction $x$, $-1\leq x\leq 1$,  parametrizes parton momenta
with respect to the symmetric momentum $P=(p+p^\prime)/2$.
In the forward limit, $p^\prime=p$, the contributions proportional to
the functions ${\cal E}^q (x,\xi,t)$ and ${\cal E}^g (x,\xi,t)$ vanish,
the distributions ${\cal H}^q (x,\xi,t)$ and ${\cal H}^g (x,\xi,t)$ reduce
to the ordinary quark and gluon densities:
\begin{eqnarray}
&&
{\cal H}^q (x,0,0)=q(x) \ \ \mbox{for} \ \ x>0 \, ,
\nonumber \\
&&
{\cal H}^q (x,0,0)=-\bar q(-x) \ \ \mbox{for} \ \ x<0 \, ;
\nonumber \\
&&
{\cal H}^g (x,0,0)=\, x \, g(x) \ \ \mbox{for} \ \ x>0 \, .
\label{reduct}
\end{eqnarray}
Note that gluon GPD is even function of $x$,
${\cal H}^g (x,\xi,t)={\cal H}^g(-x,\xi,t)$.

The meson DA $\phi_V (z)$ is defined by the following relation
\begin{eqnarray}
\label{Vdampl}
&&
\langle0|\bar q(y) \gamma_\mu q(-y)|V_L(p)\rangle_{y^2\to 0}
\\
&&
=
p_{\mu}\, f_V\!\!
\int\limits^1_0 \!dz\, e^{i(2z-1)(py)} \phi_V (z)\, .
\nonumber
\end{eqnarray}
It is normalized to unity $\int\limits^1_0 \phi_V (z) dz =1 \, $.
Here $z$ is a light-cone fraction of
a quark, $f_V$ is a meson dimensional coupling constant known from $V\to
e^+e^-$ decay, in particular, $f_\rho=198\pm 7 \, \rm{MeV}$.
\begin{eqnarray}
 {\cal M}_{\gamma^*_L N\to V_L N}
=\frac{2\pi \sqrt{4\pi\alpha}\, f_V}
{N_c \, Q \, \xi} \int\limits^1_{0} dz \, \phi_V(z)
 \int\limits^1_{-1} dx
\Biggl[   &&
\nonumber \\
\left.
 Q_V\left( T_g( z, x )\, F^g(x,\xi,t)+
T_{(+)} ( z, x)  F^{(+)} (x,\xi,t)
\right) \right. && 
\nonumber 
\end{eqnarray}
\begin{equation}
\left.
+\sum_q e_q^V T_{q} ( z, x)  F^{q(+)} (x,\xi,t)
\right] \, . \label{fact}   
\end{equation}
Here the dependence of the GPDs, DA and the hard-scattering amplitudes
on factorization scale $\mu_F$ is suppressed for shortness.
Since we consider the leading
helicity non-flip amplitude, in eq. (\ref{fact})
the hard-scattering amplitudes do not depend on $t$.
The account of this dependence would lead to the power suppressed,
$\sim t/Q$, contribution.\footnote{For the analysis of other helicity
amplitudes see \cite{MRT,IK,NNN}.}
$\alpha$ is a fine structure constant, $N_c=3$ is the number of QCD colors.
$Q_V$ depends on the meson flavor content. If one assumes it is
$\frac{1}{\sqrt{2}}(|u\bar u\rangle - |d\bar d\rangle )$,
$\frac{1}{\sqrt{2}}(|u\bar u\rangle + |d\bar
d\rangle )$ and $|s\bar s\rangle $ for $\rho$, $\omega$ and $\phi$
respectively, than $Q_\rho=\frac{1}{ \sqrt{2}}$,
$Q_\omega=\frac{1}{ 3\sqrt{2}}$ and $Q_\phi=\frac{-1}{ 3}$,
the sum in the last term of (\ref{fact}) runs over $q=u,d$ for
$V=\rho,\omega$ and $q=s$ for $V=\phi$ and
$$
e_u^\rho=e_u^\omega=\frac{2}{ 3\sqrt{2}}\, ,
\quad
e_d^\rho=-e_d^\omega=\frac{1}{ 3\sqrt{2}}\,
\quad e_s^\phi=\frac{-1}{ 3}\, .
$$
$F^{q(+)} (x,\xi,t)=
F^{q}(x,\xi,t)-F^{q}(-x,\xi,t)$ denotes a singlet quark GPD,
$F^{(+)} (x,\xi,t)=\sum_{q=u,d,s}F^{q(+)} (x,\xi,t)$ stands for the sum of
all light flavors.

Due to odd C- parity of a vector meson $\phi_V(z)=\phi_V(1-z)$. Moreover,
since $V$ and $\gamma^*$ have the same C- parities, $\gamma^*\to V$
transition selects the C-even gluon and
singlet quark contributions, whereas the
C-odd quark combination $F^{q(-)} (x,\xi,t)=
F^{q}(x,\xi,t)+F^{q}(-x,\xi,t)$ decouples in (\ref{fact}).

\noindent
{\bf 3.} Below we present the results of our calculation of the
hard-scattering amplitudes in the
$\overline{\rm{MS}}$ scheme. For that we used the dimensional regularisation of encountered infrared and ultraviolet singularities, which requires a  proper counting of number of transverse dimensions with the dimension $d = 4 + 2\epsilon = 2 + 2(1 + \epsilon)$. For more details see \cite{Ivanov:2004vd}.

$T_q(z,x)$
may be
obtained by the following substitution from the known results for a
pion EM formfactor, see also \cite{Belitsky:2001nq},
\begin{eqnarray}
\label{qqq}
&&
T_q( z, x )=\left\{T\left(z,\frac{x+\xi}{2\xi} -i\epsilon \right)
\right.
\nonumber \\
&& \left.
-
T\left(\bar z,\frac{\xi-x}{2\xi} -i\epsilon \right) \right\}
+ \Bigl\{z\to \bar z \Bigr\} \, , 
\end{eqnarray}
\begin{eqnarray}
\label{PiEM}
&&
T(v,u)= \frac{\alpha_S(\mu_R^2) C_F}{4vu}
\left( 1+\frac{\alpha_S(\mu_R^2)}{4\pi}\biggl[
\right.
\\
&&
c_1\left(2\left[3+\ln (vu)\right]
\ln{\left(\frac{Q^2}{\mu^2_F}\right)}+\ln^2 (vu)  \right.
\nonumber \\
&&
\left.
+
6\ln (vu)-\frac{\ln (v)}{\bar v}-\frac{\ln (u)}{\bar u}-\frac{28}{3}\right) 
\nonumber \\
&&
+
\beta_0\left(\frac{5}{3}-\ln (vu)-\ln{\left(\frac{Q^2}{\mu^2_R}\right)}
\right)
\nonumber \\
&&
+ c_2
\left(
2\frac{(\bar v v^2+\bar u u^2)}{(v-u)^3}
\left[
Li_2(\bar u)-Li_2(\bar v) \right.\right. 
\nonumber \\
&&
\left.
+Li_2(v)-Li_2(u)
+\ln(\bar v)\ln(u)-\ln(\bar
u)\ln(v)
\right]   
\nonumber \\
&&
+2\frac{(v+u-2vu)\ln{\bar v\bar u}}{(v-u)^2}
+2\left[
Li_2(\bar u)+Li_2(\bar v) 
\right. 
\nonumber \\
&&
\left.
-Li_2(u)-Li_2(v)+\ln(\bar v)\ln(u)
+\ln(\bar u)\ln(v) \right]
\nonumber \\
&&
\left.\left.\left.
+4\frac{vu\ln (vu)}{(v-u)^2}
-4\ln(\bar v)\ln(\bar u) -\frac{20}{3}
\right)
\right]
\right) \, . \nonumber
\end{eqnarray}
Here and below we use a shorthand notation $\bar u=1-u$ for any light-cone
fraction. $\mu_R$ is a renormalization scale for a strong coupling,
$\beta_0=\frac{11N_c}{3}-\frac{2n_f}{3}$, $n_f$ is the effective number of
quark flavors. $C_F=\frac{N_c^2-1}{2N_c}$,
$Li_2(z)=-\int\limits^z_0\frac{dt}{t}\ln(1-t)$. Also we denote
\begin{equation}
c_1=C_F \, , \quad c_2=C_F-\frac{C_A}{2}=-\frac{1}{2N_c}\, .
\label{colconst}
\end{equation}

$T_{(+)}( z, x )$ starts from NLO
\begin{equation}
\label{q+}
T_{(+)}( z, x )=\frac{\alpha_S^2(\mu_R^2)\, C_F}{(8\pi) z\bar z}\, {\cal
I}_q\left(z,\frac{x-\xi}{2\xi}+i\epsilon\right) \ ,
\end{equation}
here
\begin{eqnarray}
\label{quarktogluon}
&& {\cal I}_q(z,y)  =
\left\{
\frac{2y+1}{y(y+1)}
\left[
\frac{y}{2}\ln^2(-y)\right.\right.
\nonumber \\
&& 
-\frac{y+1}{2}
\ln^2(y+1) 
\nonumber \\
&&
+
\ln\biggl(\frac{Q^2 z}{\mu_F^2}\biggr) 
\Bigl(y\ln(-y)
\nonumber \\
&&
\left.
-(y+1)\ln(y+1)
\Bigr)\right]-\frac{R(z,y)}{y+z}
\nonumber \\
&& +
\frac{y\ln(-y)+(y+1)\ln(y+1)}{y(y+1)} 
\nonumber \\
&& \left. 
+\frac{y(y+1)+(y+z)^2}{(y+z)^2}H(z,y)
\right\}  
\nonumber \\
&&
+\Bigl\{z\to \bar
z\Bigr\} \, ,
\end{eqnarray}
where we introduced two auxiliary functions
\begin{eqnarray}
\label{R}
&
R(z,y)=z \ln (-y) +
   \bar z\ln (y+1) &
\nonumber \\
&
+ z\ln (z) +
  \bar z \ln (\bar z) \, , &
\end{eqnarray}
\begin{eqnarray}
\label{H}
&
H(z,y)=
 Li_2(y+1)-Li_2(-y)+Li_2(z) &
\nonumber \\
&
-Li_2(\bar z)
+ \ln (-y)\,\ln (\bar z) -
  \ln (y+1)\,\ln (z) \, . &
\end{eqnarray}

For the gluonic contribution we obtain
\begin{eqnarray}
\label{gluon}
&&
T_{g}( z, x )=\frac{\alpha_S(\mu_R^2) \, \xi}{z\bar z \,
(x+\xi-i\epsilon)(x-\xi+i\epsilon)}
\biggl[ 
\nonumber \\
&&
1 +
\frac{\alpha_S(\mu_R^2)}{4\pi}\, {\cal
I}_g\left(z,\frac{x-\xi}{2\xi}+i\epsilon\right)\biggr] \, , 
\end{eqnarray}
where
\begin{eqnarray}
\label{Ig}
&&
{\cal I}_g(z,y)=\left\{
\left(\ln\biggl(\frac{Q^2}{\mu_F^2}\biggr)-1\right)
\Bigl[
 \right.
\nonumber \\
&&
\frac{\beta_0}{2}-\frac{2(c_1-c_2)
(y^2+(y+1)^2)}{ y(y+1)}\Bigl(
\nonumber \\
&&
(y+1)\ln(y+1)-y\ln(-y)\Bigr)
\nonumber \\
&&
+\frac{c_1}{2}
\left(\frac{y\ln(-y)}{y+1}+\frac{(y+1)\ln(y+1)}{y} \right)
\nonumber \\
&&
\left. +c_1\left(
\frac{3}{2}+2z\ln(\bar z)\right)
 \right] -2c_1
\nonumber \\
&&
-\frac{\beta_0}{2}\left(\ln\biggl(\frac{Q^2}{\mu_R^2}\biggr)-1\right)
-\frac{c_1(2y+1)R(z,y)}{2(y+z)} 
\nonumber \\
&&
-\frac{(3c_1-4c_2)}{4}\left(
\frac{y\ln^2(-y)}{y+1}\right.
\nonumber \\
&&
\left.
+\frac{(y+1)\ln^2(y+1)}{y}\right)+c_1(1+3z)\ln(\bar z)
\nonumber \\
&&
+\Bigl(\ln(-y)+\ln(y+1)\Bigr)\Bigl[
\nonumber \\
&&
\left. 
c_1\left(\bar z\ln(z)-\frac{1}{4}\right)
+2c_2\right] 
\nonumber 
\end{eqnarray}
\vspace*{-0.6cm}
\begin{eqnarray}
&&
-(c_1-c_2)\Bigl(\ln(z\bar z)-2\Bigr)\!
\left[\frac{y\ln(-y)}{y+1}\right.
\nonumber \\
&&
\left. 
+\frac{(y+1)\ln(y+1)}{y}
\right] + c_1 z\ln^2(\bar z)
\nonumber \\
&&
+
(c_1-c_2)(2y+1)\ln\left(\frac{-y}{y+1}\right)
\nonumber \\
&&
\left[\frac{3}{2}+\ln(z\bar z)+\ln(-y)+\ln(y+1)\right]
\nonumber \\
&&
+\Bigl(c_1(y(y+1)+(y+z)^2)
\nonumber \\
&&
-c_2(2y+1)(y+z)\Bigr)\biggl[-\frac{R(z,y)}{(y+z)^2}
\nonumber \\
&&
\left.
+\frac{\ln(-y)-\ln(y+1)+\ln(z)-\ln(\bar z)}{2(y+z)}
    \right.
\nonumber \\
&&
\left.
+\frac{y(y+1)+(y+z)^2}{(y+z)^3}\, H(z,y)\right]
\biggr\}
\nonumber \\
&&
+\Bigl\{z\to \bar z\Bigr\} \, .
\label{Igluon}
\end{eqnarray}

\noindent
{\bf 4.} Above formulas and the known NLO evolution
equations describing $\mu_F$ dependence of the GPDs and meson
DA give a complete basis for the description of a neutral
vector meson electroproduction with NLO accuracy.
At leading order we reproduce the
known result \cite{Mank97}, our results for the NLO hard-scattering
amplitudes are new.

At high energies, $W^2\gg Q^2$, the imaginary part of the amplitude
dominates. Leading contribution to the NLO correction comes from the
integration region $\xi\ll |x|\ll 1$, simplifying
the gluon hard-scattering amplitude in this limit we obtain the estimate
\begin{equation}
 {\cal M}_{\gamma^*_L N\to V_L N} 
\approx  \label{appr} 
\end{equation}
\vspace*{-0.3cm}
\begin{eqnarray}
 \frac{-2\, i\, \pi^2 \sqrt{4\pi\alpha}\, \alpha_S f_V Q_V}
{N_c \, Q \, \xi}\int\limits^1_{0}\frac{ dz \, \phi_V(z)}{z\bar z}\Biggl[
F^g(\xi,\xi,t)  &&
\nonumber \\
+\frac{\alpha_S N_c}{\pi}
\ln\left(\frac{Q^2z\bar z}{\mu_F^2}\right)
\int\limits^1_{\xi} \frac{dx}{x}  F^g(x,\xi,t)
\Biggr] \, . \nonumber &&
\end{eqnarray}
Given the behavior of the gluon GPD at small $x$, $F^g(x,\xi,t)\sim const$,
we see that NLO correction is parametrically
large, $\sim \ln(1/\xi)$, and negative unless one chooses the
value of the factorization scale sufficiently lower than the kinematic
scale. For the asymptotic form of meson DA, $\phi^{as}_V(z)=6z\bar z$,
the last term in (\ref{appr}) changes the sign at $\mu_F=\frac{Q}{e}$,
for the DA with a more broad shape this happens at even lower values of
$\mu_F$.

It is interesting to note that the value $\mu_F^2=Q^2/e^2$
is rather close to an estimate in the dipole approach
\cite{NNN0} of a typical inverse dipole size  \cite{NNN1,NNN2},
$1/r^2\sim \mu_F^2=0.15 Q^2$, for vector meson
electroproduction in the HERA kinematic region.
We believe that
a study of relationship between the collinear factorization and
the dipole approach at the NLO level would be very important. But this
question, as well as the resummation of large at high energies,
$\sim (\alpha_S\ln(1/\xi))^n$, contributions both to the hard-scattering
amplitudes
and to the evolution of GPDs, go beyond the scope of the present work.

\begin{figure}[t]
\epsfxsize7.5cm\epsffile{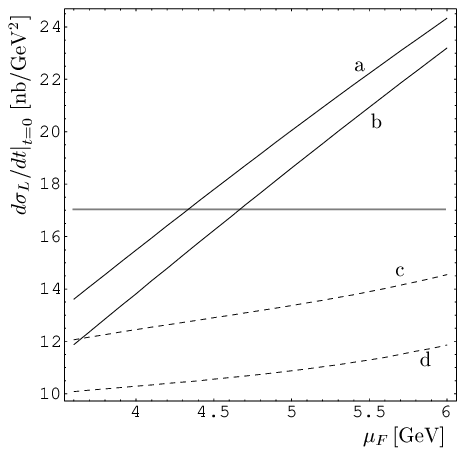}
\caption{
Factorization scale dependence of predicted 
$ d \sigma_L/dt|_{t=0}$ at $Q^2=27\,\,{\rm GeV^2}$, 
$W=110\,\,{\rm GeV}$.
The horizontal black line describes the data point. 
Solid curves assume $\mu_R=\mu_F$, 
dashed curves -- the BLM prescription.
The curves a and c (b and d) are obtained with the use of 
MRST2001 (CTEQ6M).
}
\label{fig:1}
\end{figure}

\noindent
{\bf 5.} As an example of our results we compare
a one point for the longitudinal cross section
reported by ZEUS collaboration  
\cite{Breitweg:1998nh}, e.g. 
$ d \sigma_L/dt|_{t=0}=17\pm 4 \,\,{\rm nb/GeV^2}$ at 
$Q^2=27\,\,{\rm GeV^2}$, $W=110\,\,{\rm GeV}$, with our predictions.
On the Fig.~1 we plot the dependence of predicted 
$ d\sigma_L/dt|_{t=0}(\mu_F,\mu_R)$ on the factorization scale $\mu_F$
for two choices of the renormalization scale: 
$\mu_R=\mu_F$ (the solid curves) and $\mu_R=Q/\sqrt{e}$, i.e. 
the BLM (Brodsky-Lepage-McKenzie) prescription (the dashed curves). 
The data point is described in this plot by the
black horizontal line. In this numerical analysis we use two models
of the NLO generalized parton distributions of \cite{Freund:2002qf}:
first one 
based on MRST2001 forward distribution
(the curves a and c) and second one -- on
CTEQ6M (the curves b and d). 
Moreover, we take the NLO strong coupling constant $\alpha_s$ and 
the asymptotic meson DA\footnote{
QCD sum rules studies \cite{Braun}
show that at low scale the shape of vector meson
DA is close to the asymptotic one. At $\mu_F\to \infty$ any DA
aproaches the limit -- asymptotic DA.
Due to this we use asymptotic DA, and postponed the study of dependence on
the DA shape for the future analysis.}.
Fig.~1 shows that
the BLM prescription leads to smaller cross section and to
much flatter behaviour of the
cross-section on $\mu_F$ than for the choice $\mu_R=\mu_F$. We observe also
a substantial uncertainty of our predictions due to the input parton GPDs.

Since as mentioned above the NLO corrections are large it is
instructive to study the relative magnitudes of 
different contributions to the amplitude (10).
This is done by assuming  $\mu_R=\mu_F$ and for CTEQ6M GPD. 
In Fig.~2 we show
plots of  $Im {\cal M}$ and $Re {\cal M}$ as functions of $\mu_F$ 
corresponding to a gluonic and
a quark Born parts of the amplitude (10)
(denoted as $gB$ and $qB$, respectively), 
to the NLO part of the gluonic 
contribution $T_g$ ($gN$), to the NLO part of the quark
contribution $T_q$ ($qN$), to the quark contribution
$T_{(+)}$ ($q^+$) and to the full amplitude (10) (denoted as full). All
these
separate contributions are normalized by $|{\cal M}|$ of (10) with all
terms taken into account. Let us note that, as expected for the 
small $x$ process, 
the imaginary part of the amplitude (10) dominates over its real part.
We observe that the NLO corrections are mostly of opposite signs than
the corresponding Born terms and they are big,  consequently 
the final values of the
amplitude (10) are the result of a strong cancellations between
Born parts and NLO terms. Without account of NLO terms the
predictions would be substantially above the data.
These results are similar to ones obtained
recently for $\Upsilon$ photoproduction \cite{UPS}.

In this paper we restrict our analysis to the leading twist and neglect
completely the power suppressed, $\sim 1/Q$, contributions. This is
definitely legitimate for sufficiently large $Q$. Most probably, at the HERA
energies and $Q^2\sim 20\, \rm{GeV}^2$ the higher twist corrections
are still large, see e.g. \cite{NNN1,NNN2} based on \cite{Shur}.
Nevertheless, we want to stress
that our leading twist results obtained with NLO
hard-scattering amplitudes and NLO GPDs \cite{Freund:2002qf} (
which were adjusted to describe
HERA deeply virtual
Compton scattering data)
are in qualitative agreement with the measured at HERA $\rho$ meson
electroproduction cross section.

\vspace*{1cm}

\centerline{
{\bf  Acknowledgments}}

\vspace*{0.5cm}

\noindent Work of D.I. is supported in part by Alexander von
Humboldt Foundation and by grants DFG 436, RFBR 03-02-17734, L.Sz.
is partially supported by the French-Polish scientific agreement
Polonium.

\begin{figure}[t]
\epsfxsize7.5cm\epsffile{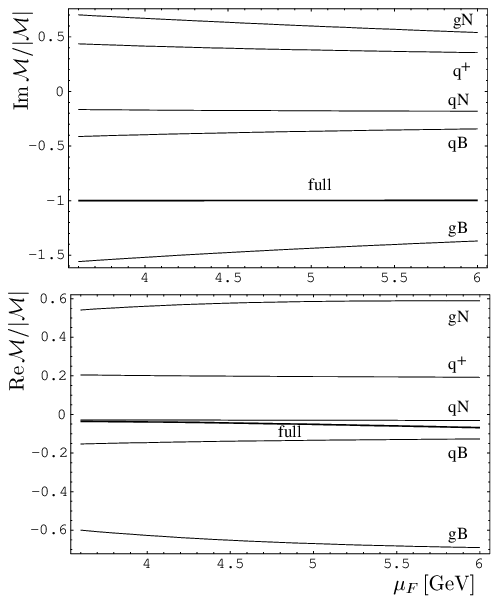}
\caption{
Different contributions
to
$Im {\cal M} /|{\cal M}|$ and $Re {\cal M}/ |{\cal M}|$
in dependence on
$\mu_F$
(for $\mu_R=\mu_F$ and CTEQ6M): 
$gB$ -- Born term of $T_g$, $qB$ -- Born term of $T_q$, 
$gN$ -- NLO terms of $T_g$, $qN$ -- NLO terms  of $T_q$, 
$q^+$ -- $T_{(+)}$, full -- all terms of (10) included.
}
\label{fig:2}
\end{figure}

\end{document}